\documentclass[aps,prb,twocolumn,superscriptaddress]{revtex4-2}

\usepackage{amsmath,amsfonts}
\usepackage{graphicx}
\usepackage[colorlinks=true, allcolors=blue]{hyperref}
\usepackage{color}

\begin{document}

\title{Time-resolved dynamics of GaN waveguide polaritons}

\author{Loïc Méchin}\email[]{loic.mechin@uca.fr}
\affiliation{Université Clermont Auvergne, Clermont Auvergne INP, CNRS, Institut Pascal, F-63000 Clermont-Ferrand, France}
\author{François Médard}
\affiliation{Université Clermont Auvergne, Clermont Auvergne INP, CNRS, Institut Pascal, F-63000 Clermont-Ferrand, France}
\author{Joël Leymarie}
\affiliation{Université Clermont Auvergne, Clermont Auvergne INP, CNRS, Institut Pascal, F-63000 Clermont-Ferrand, France}
\author{Sophie Bouchoule}
\affiliation{Centre de Nanosciences et de Nanotechnologies, CNRS, Université Paris-Saclay, F-91120 Palaiseau, France}
\author{Blandine Alloing}
\affiliation{Université Côte d’Azur, CNRS, CRHEA, rue Bernard Gregory, Sophia Antipolis, F-06560 Valbonne, France}
\author{Jes\'us Z\'uñiga-Pérez}
\affiliation{Université Côte d’Azur, CNRS, CRHEA, rue Bernard Gregory, Sophia Antipolis, F-06560 Valbonne, France}
\affiliation{Majulab, International Research Laboratory IRL 3654, CNRS, Université Côte d’Azur, Sorbonne Université, National University of Singapore, Nanyang Technological University, Singapore 117543, Singapore}
\author{Pierre Disseix}
\affiliation{Université Clermont Auvergne, Clermont Auvergne INP, CNRS, Institut Pascal, F-63000 Clermont-Ferrand, France}

\date{\today}

\begin{abstract}
We implement a new experimental approach to directly measure the lifetime of guided polaritons arising from the strong-coupling of GaN excitons and the guided photonic modes of a slab waveguide. Using a Fourier imaging setup, combined with spatial filtering of the emission, the emission associated to polaritonic modes with well-defined propagation constants can be selectively analyzed in the temporal domain. By directing it to the entrance slit of a streak camera, time-resolved photoluminescence (TRPL) measurements along the polariton dispersion branch were performed at 40~K, enabling to assess the time decay of polariton modes. By combining this information with the photonic/excitonic fraction corresponding to each polariton mode, extracted from a coupled-oscillators model that indicate a Rabi splitting of $\Omega$~=~80~meV, we could extract the photon lifetime in the waveguide $\tau_\gamma\, =\, 3\pm 1$~ps. This corresponds to a record $Q$-factor in the UV of 16~000. The excitonic reservoir lifetime, which contributes to polariton formation, was determined through TRPL measurements on excitonic luminescence. Finally, measurements conducted at lower temperature highlight secondary feeding mechanisms for the guided polaritonic mode, either via photon recycling from the AlGaN cladding layer or through resonant injection of photons from transitions below the band gap.
\end{abstract}


\maketitle

\section{Introduction}

In semiconductor materials, the reversible exchange of energy between electromagnetic modes and excitonic resonances leads to the formation of hybrid light-matter quasiparticles known as exciton-polaritons or simply polaritons. Initially theorized by Hopfield and Agranovich in the late 1950s \citep{Hopfield_1958, Agranovitch_1959}, advances in the fabrication of submillimeter heterostructures such as planar microcavities \cite{Weisbuch_1992} and waveguides \cite{Walker_2013} have enabled the exploration of various bosonic properties of polaritons. These include the Bose-Einstein condensation in open systems \citep{Kasprzak_2006, Kasprzak_2008, Richard_2005, Sun_2017}, superfluidity of light \citep{Sanvitto_2016, Amo_2009}, the study of topological defects \citep{Lagoudakis_2009, Lagoudakis_2008, Hivet_2012}, and the generation of spin flows \citep{Kavokin_2005, Leyder_2007}. These studies have led to the observation of numerous phenomena in these quantum fluids of light, such as parametric amplification and oscillation \citep{Savvidis_2000, Saba_2001, Stevenson_2000}, soliton formation \cite{Walker_2015}, and coherent light emission without the need for population inversion \cite{Imamoglu_1996}. Collectively, these works have demonstrated the feasibility of achieving interaction between two light beams if dressed with some matter component, opening the way for the development of all-optical, low-power and micrometer-sized devices.\par
The planar microcavity geometry has allowed to attain the strong-coupling regime in wide bandgap semiconductors and, subsequently, the Bose-Einstein condensation of polaritons up to room-temperature, establishing thereby the state-of-the-art for this family of semiconductors. However, the need for Distributed Bragg Reflectors (DBRs) complexifies their fabrication, restricts the attainable photonic lifetime ($Q$-factor), and limits their exploitation in polaritonic devices where several functionalities could be concatenated. The use of waveguide-type geometry offers advantages for the realization of such devices: (i) the confinement of optical modes propagating in the waveguide through total internal reflection ensures strong overlap with the excitonic resonances, enabling the strong coupling regime with excitons in the active material \cite{Brimont_2020}; (ii) the development of electrical carrier injection is much simpler in waveguide geometry due to the simplicity of the heterostructure \citep{Rosenberg_2016, Suarez_2020}; (iii) the high group velocity of the guided modes (on the order of $10^7 \, \mathrm{m\cdot s^{-1}}$) gives access to fast-propagating polaritons, which are necessary for high-frequency device applications \citep{Walker_2015, Amo_2016, Leblanc_2020}. \par
To operate at room-temperature, polaritonic waveguides must be based on semiconductors displaying large binding energies and large oscillator strength. Commonly studied platforms include organic systems \citep{Zasedatelev_2019, Ellenbogen_2011, Pirotta_2014} or inorganic systems based on ZnO \citep{Jamadi_2017, Kreyder_2023}, chosen for their large Rabi splitting. GaN-based platforms emerge as promising candidates due to (i) the large binding energy of excitons, which ensures the stability of polaritons at room temperature \citep{Christopoulos_2007, Christmann_2008, Di-Paola_2021}, similar to ZnO, and (ii) well-established technological processes such as doping \citep{Amano_1989, Nakamura_1992} and clean-room processing \cite{Souissi_2022}, which are not accessible in other semiconductor platforms, as ZnO.\par 
In the case of waveguide polariton lasers, besides the standard polariton relaxation mechanisms operating in vertical microcavities, as polariton-polariton and polariton-phonon interactions \citep{Tassone_1997, Tassone_1999}, one needs to take into account the scape of polaritons below the pumping spot due to the generation of polaritons with large in-plane velocities \cite{Solnyshkov_2014}. Recently, it has been demonstrated that GaN waveguide polaritons spatially-confined by vertical DBRs, can provide enough gain to achieve polariton lasing in such waveguides, and this up to room-temperature \citep{Souissi_2022, Souissi_2024}. To further reduce the lasing threshold and develop low-power consumption polaritonic devices, unravelling the physical mechanisms underlying this polaritonic gain is necessary.\par 
To determine the excitonic and photonic fractions of waveguided polaritons one can relie on the experimental measurement of the dispersion curve and its subsequent numerical simulation to extract the associated Hopfield coefficients for each state \cite{Brimont_2020}, similar to vertical microcavities. On the other hand, because the photonic lifetime of the waveguides can be hardly measured, contrary to vertical cavities where $Q$-factor measurements are routinely made, one can not easily access or extrapolate the polaritons lifetime along their dispersion. In this context, Ciers et al. proposed estimating polariton lifetimes by measuring their diffusion lengths, combined with the knowledge of their group velocity along the dispersion curve \citep{Ciers_2017}. In this work, we present an alternative experimental method that uses time-resolved photoluminescence to measure the lifetime of each state along the polaritonic dispersion branch. This technique not only enables the determination of polariton lifetimes along the dispersion curve but also provides insight into the physical mechanisms feeding the polaritonic modes in the structure.

\section{Experimental details}

\subsection{Sample description}

\begin{figure}
	\includegraphics[width=\columnwidth]{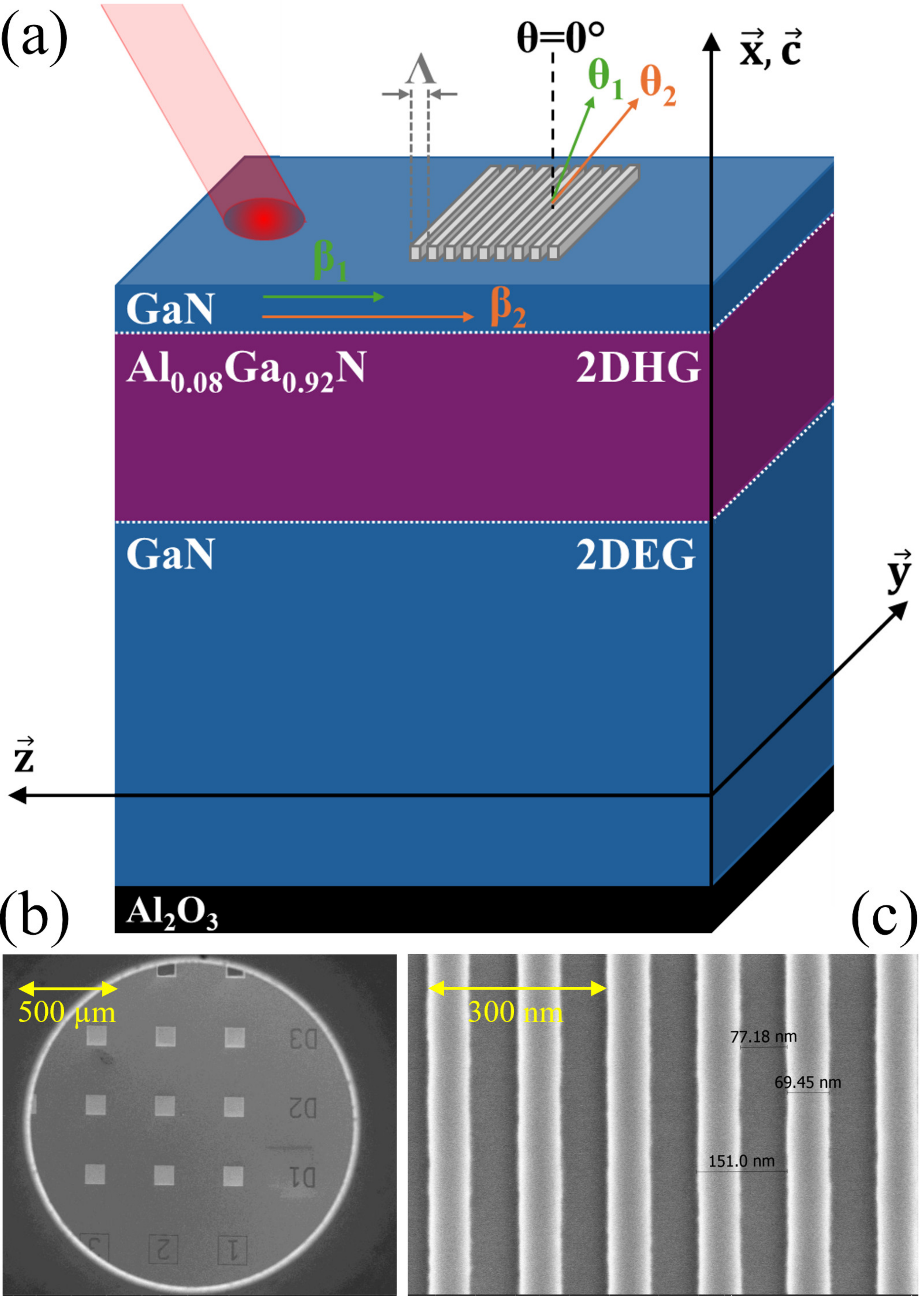}
	\caption{\label{Fig1} (a) Schematic representation of the studied sample. Blue regions correspond to GaN layers (150 nm and 3.5 µm). Purple region represents AlGaN lower cladding (1.5 µm) and black region displays the sapphire substrate (thicknesses are not at scale). The excitation laser beam and a grating are also representated. Two guided modes with different in-plane wavevectors are extracted by the grating with different emission angle to illustrate the Bragg's diffraction law (\textcolor{blue}{eq.}\ref{eq:Bragglaw}). (b) Scanning electron microscope (SEM) image showing the gratings elaborated by electron beam lithography on top of the structure. (c) Scanning electron microscope (SEM) image showing grating lines, the good periodicity prevents the detection of ghost lines in angle-resolved photoluminescence measurements.}
\end{figure}

The structure investigated in this work consists of a GaN optical waveguide (150 nm thick) with an AlGaN lower cladding (1.5 µm thick) grown onto a $c$-plane sapphire substrate with a GaN buffer layer (3.5 µm) by metal-organic chemical vapor deposition (MOCVD). A schematic representation of the structure is displayed in Fig. \ref{Fig1}\textcolor{blue}{(a)}. A state-of-the-art dislocation density of $3\times 10^8 \mathrm{cm^{-2}}$ is obtained by starting the growth with a 3D mode that is followed by a 2D growth to obtain a flat surface with low roughness ($\sim$ 1 nm for $5\times 5 \, \mathrm{\mu m^2}$). Upon cooling, a uniform stress is distributed across the structure, due to the difference between sapphire and nitrides coefficients of thermal expansion \cite{Detchprohm_1992}. The growth is performed along the [0001] metal-polarity direction. This leads to a presence of a two-dimensional hole gas (2DHG) and a two-dimensional electron gas (2DEG) at the GaN/AlGaN and AlGaN/GaN interfaces respectively, as represented in Fig. \ref{Fig1}\textcolor{blue}{(a)}. The 2DHG and the effects of polarization field in this sample have been observed and characterized through continuous and time-resolved spectroscopy in previous works \citep{Mechin_2024, Mechin_2025}.\par
The thickness of the GaN guiding layer (150 nm) and the low aluminium composition (8\%) of the AlGaN lower cladding were chosen to support the propagation of the fundamental mode $\mathrm{TE_0}$ of the waveguide with a good confinement while allowing pseudomorphic growth to be maintained, minimizing the incorporation of stress and defects in the surface layer. As displayed on the scanning electron microscope (SEM) image in Fig.\ref{Fig1}\textcolor{blue}{(b)}, $\mathrm{SiO_2}$ gratings ($100\times 100\, \mathrm{\mu m^2}$) were deposited by electron beam lithography on top of the structure to enable the outcoupling of the guided modes that lie below the light-cone. Different electron doses and different periods ($\Lambda = \left[ 134, 140, 146, 152, 158, 162\right]$ nm) were used to ensure a good extraction and analysis of the guided modes. As shown in Fig. \ref{Fig1}\textcolor{blue}{(c)}, the periodicity of the grating lines is achieved with high precision ($<$ 2 nm), preventing the detection of ghost lines in angle-resolved photoluminescence measurements.

\subsection{Experimental setup}

\begin{figure}
	\includegraphics[width=\columnwidth]{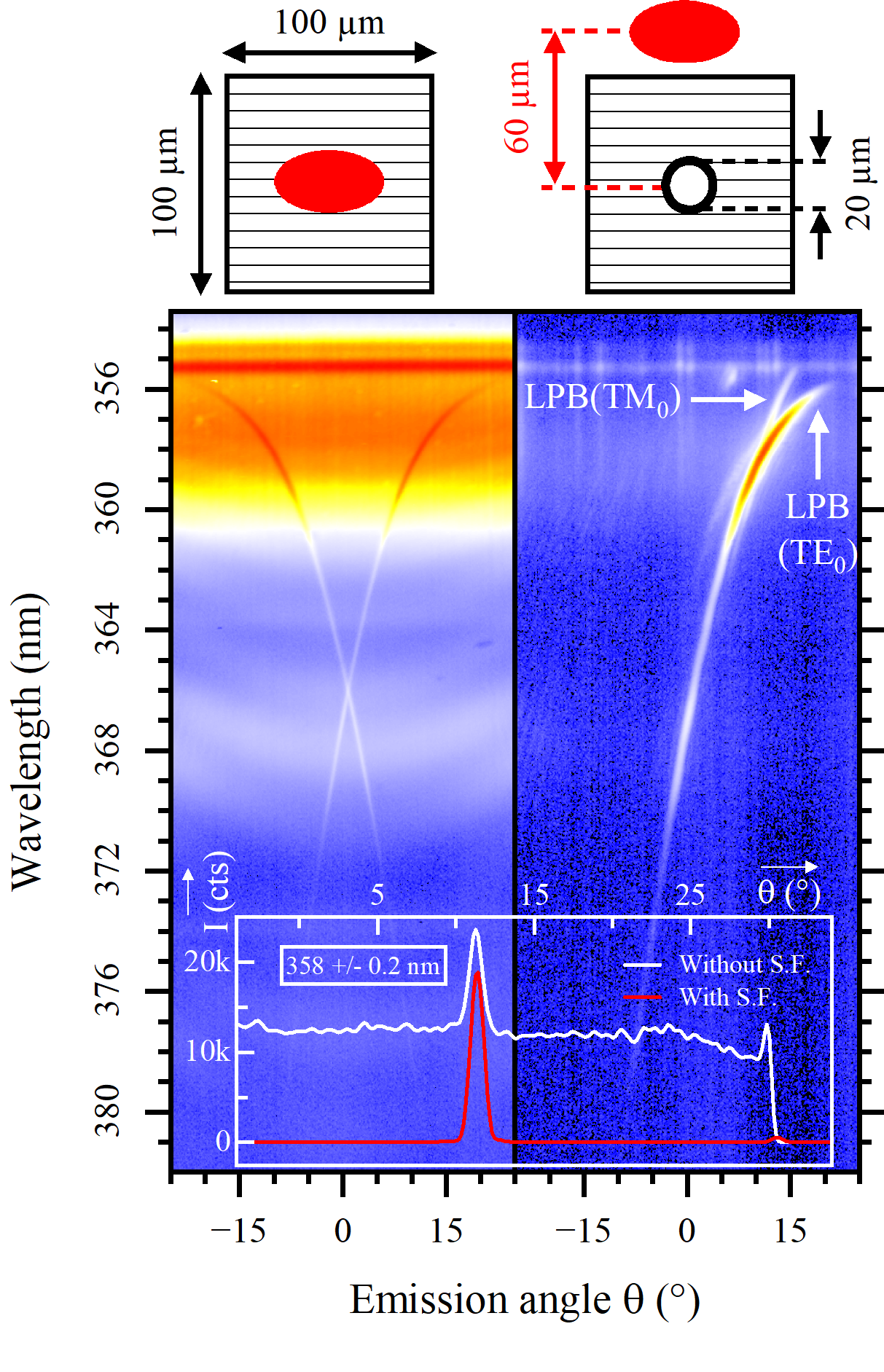}
	\caption{\label{Fig2} Left panel: Measurement of the polariton dispersion curve under the excitation spot without spatial filtering (S.F.). Right panel: Measurement of the dispersion curve after shifting the excitation spot above the extraction grating and using a pinhole to apply spatial filtering, analyzing only a 20 µm-diameter region at the center of the extraction grating. In this configuration, only one propagation direction of the polaritons is collected, and nearly all of the sample's direct luminescence is filtered out, as illustrated by the inset white graph. This graph displays the integrated intensity over both images within a wavelength range of 358.0 ± 0.2 nm.}
\end{figure}

The sample was placed in a closed-cycle helium cryostat, enabling the study of its luminescence properties over a wide temperature range (from 3 K to room temperature). The optical setup used to perform continuous and time-resolved luminescence measurements is equipped with a Ti:sapphire laser with a pulse duration of 150 fs and a repetition rate of 76 MHz. The combined use of a Lyot filter and a nonlinear crystal enables access to a broad range of excitation wavelengths. In this work we have used the second and the third harmonics, at 349 nm and 266 nm respectively, of the laser output wavelength. Then, the laser beam is focused on the sample surface using an aspherical lens with an angle of 64° to avoid the reflection of the laser signal into the detection line. The emission of the sample was collected through a 100 $\times$ NUV Mitutoyo microscope objective with a 0.5 numerical aperture and 2 mm focal length. The use of a spherical lens with a focal length of 200 mm allows to image the surface of the sample on a 2048$\times$512 pixels CCD (Charged-Coupled Device) camera with a magnification of 100. A spectrometer equiped with a 600 grooves per mm grating is used to spectrally disperse the light and obtain µPL spectra. Time-resolved measurements were realized with a Hamamatsu streak camera placed on the output plane of the spectrometer with a maximal resolution of about 3 ps.\par
Figure \ref{Fig2} displays measurements of the dispersion of guided polaritons obtained on the CCD camera. These results were acquired by adding a lens of 75 mm focal length to image the Fourier plane onto the entrance slit of the spectrometer. Bragg’s diffraction law allows us to relate the emission angle $\theta$ to the propagation constant of the mode $\beta$:
\begin{equation}\label{eq:Bragglaw}
\beta =2\pi\left[ \frac{\sin (\theta )}{\lambda_0} - \frac{\mathrm{m}}{\Lambda}\right] \quad \mathrm{with} \, \mathrm{m}\in \mathbb{Z}.
\end{equation}\par
To perform time-resolved analysis of each state along the polariton dispersion curve, the Fourier plane is imaged on the entrance slit of the streak camera. For the measurements presented in this work the slit width was set to 50 µm, corresponding to an angular resolution of 0.5°. The collection lens can be moved vertically thanks to a micrometric screw allowing a vertical translation of the Fourier plane. It enables a time-resolved analysis of the states along the dispersion curve as a function of the emission angle $\theta$ (or the propagation constant $\beta$).\par
Two excitation and detection configurations were tested to improve the signal over noise ratio for the polaritonic dispersion. In the first one, shown on the left side of Figure \ref{Fig2}, the laser spot is placed directly on the extraction grating and the signal is collected at the same place. This approach has the advantage of providing a stronger signal. However, time-resolved analysis is challenging because the unguided luminescence of the sample overlaid the guided mode emission.\par
To address this issue, a second configuration using real-space filtering was employed. As shown on the right side of Figure \ref{Fig2}, by exciting the sample beside the extraction grating and using a pinhole with a diameter of 0.8 mm, we can detect only the light emitted from the guided modes extracted by the grating over a region of approximately 20 µm in diameter. The inset in Figure \ref{Fig2} demonstrates that spatial filtering effectively eliminates nearly all of the unguided photoluminescence from the sample, making the polariton dispersion curve accessible for time-resolved analysis. 

\section{Direct measurement of polaritonic lifetime along the dispersion curve}
\begin{figure*}
\includegraphics[width=175mm]{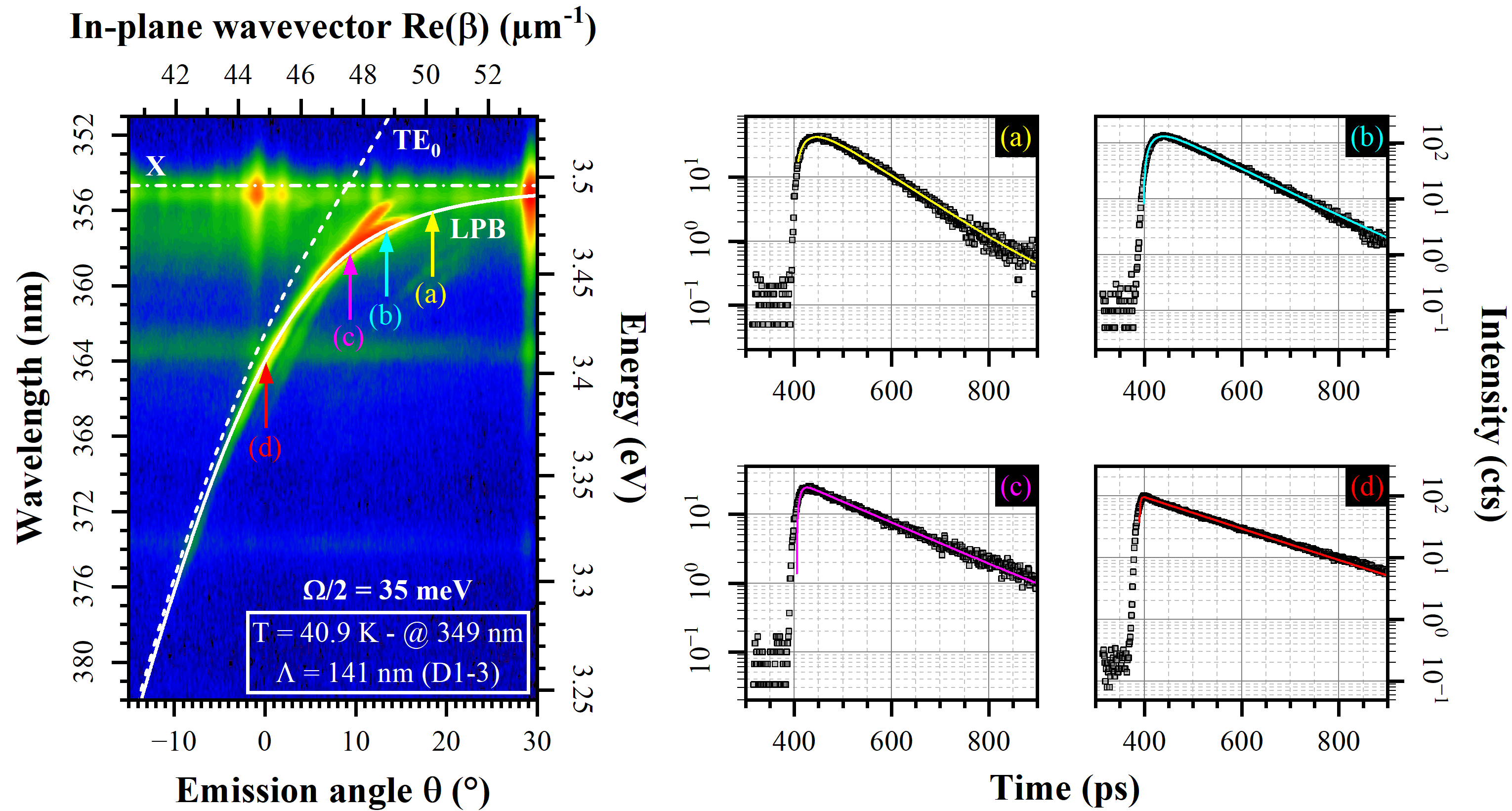}
\caption{\label{Fig3} Left panel: experimental measurement of the lower polariton branch (LPB) for a temperature of T = 40.9 K and a laser excitation power of 10 mW at 349 nm (the color scale corresponds to a logarithmic intensity scale). The solid white line represents the theoretical calculation of the polaritonic dispersion using a quasiparticle model. The dashed white lines correspond to the dispersion of the photonic mode propagating in the waveguide ($\mathrm{TE_0}$) and the excitonic resonance in the material at 3.4948 eV. The second mode observable above 3.46 eV, displaying a lower effective mass, corresponds to the $\mathrm{TM_0}$ mode strongly-coupled to GaN C exciton, as confirmed by polarization-resolved dispersion measurements. Right panel: temporal decay of polaritonic states (a), (b), (c), and (d) highlighted in the left panel. The gray squares represent the experimental data, and the colored lines correspond to fits of the decays using equation \ref{eq4}.}
\end{figure*}
The lifetime measurements of polaritons along the dispersion curve were performed at a temperature of 40.9 K. The laser excitation wavelength and power were set to 349 nm and 10 mW, respectively. Figure \ref{Fig3} shows the dispersion curve of the guided polaritonic mode under these experimental conditions as a false-color image with a logarithmic scale for intensity (left panel). Both $\mathrm{TE_0}$ and $\mathrm{TM_0}$ modes are visible and superimposed for wavelengths longer than about 360~nm, as observed in Figure~\ref{Fig2}, where each mode is highlighted. The $\mathrm{TE_0}$ mode is strongly coupled with the A and B excitons, while the $\mathrm{TM_0}$ mode is coupled at higher energy with the C exciton, in accordance with the selection rules in GaN and confirmed by polarization-resolved dispersion measurements. However, we will focus only on the $\mathrm{TE_0}$ mode whose intensity is higher. We can also notice that the photoluminescence of excitons (both free and bound) and their longitudinal optical phonon replica are not completely removed by spatial filtering and can be seen as non-dispersive signals aroud 355 nm and 363.5 nm. Time-resolved measurements were performed for emission angles $\theta$ ranging from -11.4° to 17.8°. Four of these measurements are displayed (right panel) for angles of 17.8°, 13.5°, 9.2°, 0.0°, corresponding to points (a), (b), (c), and (d) indicated in the polariton dispersion. For each measured angle the photoluminescence intensity decreases, for sufficiently long times, with the same monoexponential decay (time decay~$\sim$~130~ps); meanwhile, the rising time of the photoluminescence intensity at short delay times varies from a very short one at large wavelengths (close to the setup resolution), to longer rising times as the emission wavelength approaches the excitons resonance. To extract the polariton lifetime from these experimental time-resolved measurements, we have modeled the temporal evolution of the polariton population in the waveguide. During laser excitation, charge carriers are created at high energy. After thermalization, they populate a reservoir of uncoupled excitons. Exciton-exciton and exciton-phonon interactions within the reservoir allow further relaxation, enabling them to couple with the photonic mode of the waveguide, forming polaritons. The temporal dynamics of the excitonic and polaritonic populations can be expressed as:
\begin{align}
\frac{dn_X (t)}{dt}&=G(t) - \frac{n_X (t)}{\tau_X};\\
\frac{dn_{pol} (\beta ,t)}{dt}&=\frac{n_X (t)}{\tau_t (\beta)} - \frac{n_{pol}(\beta ,t)}{\tau_{pol}(\beta)};
\end{align}
where $n_X$ and $n_{pol}$ are the populations of excitons in the reservoir and polaritons in the state $\beta$, respectively. $G$ represents the generation term, considered as a Dirac function. $\tau_X$ is the exciton lifetime within the reservoir, while $\tau_t$ represents all relaxation paths towards the given polariton state $\beta$; thus, it is important to note that $\tau_t$ is not equal to $\tau_X$. Finally, $\tau_{pol}$ is the polariton lifetime in the state $\beta$. Solving the differential equations allows to find an analytical expression for the polariton density in the material as a function of time:
\begin{equation}\label{eq4}
n_{pol} (\beta ,t)=A\left[ \exp\left\lbrace -\frac{t}{\tau_{X}}\right\rbrace - \exp\left\lbrace -\frac{t}{\tau_{pol} (\beta)}\right\rbrace\right] ,
\end{equation}
where $A = n_X^0 \tau_X \tau_{pol} (\beta) \times \left[ \tau_t (\beta) \left( \tau_X - \tau_{pol}(\beta)\right)\right]^{-1} $ is set as fitting parameter. The fitting of experimental results with this model allowed us to experimentally determine the polariton lifetime along the dispersion curve. The decrease in in-plane wavevector (i.e. for larger photonic fraction) is accompagnied by a decrease in the polaritonic lifetime, from almost 30 ps for the largest wavevector analyzed to 2 ps for the lowest energy states. At the same time, the exciton reservoir lifetime is found to be between 100 ps and 160 ps, depending on the polariton density, as discussed in the next paragraph.\par 
\begin{figure}
	\includegraphics[width=\columnwidth]{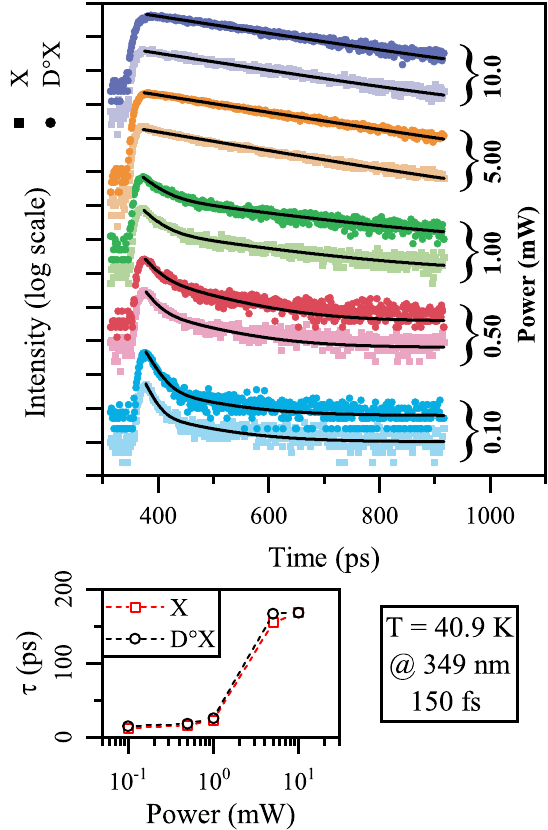}
	\caption{\label{Fig4} Top panel: time-resolved photoluminescence (TRPL) measurement as a function of excitation power for the free exciton transition X (colored squares) and the neutral donor bound exciton transition $\mathrm{D^\circ X}$ (colored circles). The black lines correspond to the fitting of these decays using a biexponential model for the three lowest excitation powers (from 100 µW to 1 mW) and a monoexponential model for the higher excitation powers (5 mW and 10 mW). Bottom panel: decay times obtained by fitting the experimental decays for the free exciton (red square) and bound exciton (black circle). Only the shortest decay time is shown, as it corresponds to the excitonic population in the thin surface GaN layer. The baseline of each spectrum is shifted vertically for clarity.}
\end{figure}
In order to determine the exciton reservoir lifetime, TRPL measurements on the sample under the same experimental conditions as those used for the polariton dispersion curve measurements (T = 40.9 K - excitation wavelength set at 349 nm) were performed. The measured temporal decay of excitonic transitions, both of free excitons and neutral donor bound excitons, as a function of excitation power is presented in Figure \ref{Fig4}. At low excitation power (from 100 µW to 1 mW), a biexponential decay of the free and bound exciton luminescence is observed. These two contributions are associated to the presence of two excitonic populations in the material, located respectively in the upper thin GaN layer (150 nm) and the lower thick GaN layer (3.5 µm) as demonstrated in a previous work \cite{Mechin_2025}. The observation of the excitonic luminescence signal from the lower GaN layer is made possible by the fact that not all photons from the excitation laser are absorbed by the top GaN nor by the AlGaN cladding, whose bandgap is larger than the excitation laser energy. The shortest decay time (12 ps) is associated with excitons from the upper GaN layer, sensitive to the built-in electric field present in the waveguide, while the longest decay time is associated with excitons from the lower GaN layer and corresponds to the decay time measured in thick GaN layer deposited on a sapphire substrate (90 ps) \cite{Mechin_2024}. At high excitation power (5 mW - 10 mW), a monoexponential decay of the excitonic transitions is observed, with a decay time longer than the one measured in bulk at 5.3 K (168 ps versus 90 ps in bulk). The transition from a very short decay time at low excitation power (12 ps at 100 µW) to a very long decay time (168 ps at 10 mW) is explained by the screening of the built-in electric field in the thin surface GaN layer by optically injected free carriers and was discussed in reference \cite{Mechin_2025}. Note that, compared to the reference \cite{Mechin_2025} where all experiments were carried out at 5.3~K, the increase in temperature in the current work (40.9~K) results in two marked differences: on the one hand the excitonic lifetime becomes larger than that in bulk GaN and, concomintantly, the free and donor bound excitons lifetimes become equal. As described by Hauswald et al. in GaN nanowires \citep{Hauswald_2014, Hauswald_2013}, at 40 K, the thermal energy is sufficient to couple the excitonic reservoir and the bound excitons, allowing delocalization of the otherwise bound excitons. This mechanism results in free and bound excitons exhibiting the same lifetime, as observed in our experimental results. These findings also explain the choice of an excitation power of 10 mW, which minimizes the effects of the electric field in the structure during TRPL measurements along the polaritonic dispersion curve.\par
Knowing the lifetime of the polariton $\tau_{pol}$ along the dispersion curve and the excitonic lifetime $\tau_{X}$, we can access the guided photon lifetime $\tau_{\gamma}$ using a two coupled oscillators model as described by Hopfield \cite{Hopfield_1958}:
\begin{align}
\frac{1}{\tau_{pol}}&=\frac{\vert X\vert^2}{\tau_X}+\frac{\vert P\vert^2}{\tau_\gamma};\label{eqtpol}\\
\mathrm{where} \quad X&=\left[ \sqrt{1+4\left(\frac{E_{LPB}(\beta)-E_X(\beta)}{\Omega}\right)^2}\right]^{-1};\\
\mathrm{and} \quad P&=\left[ \sqrt{1+\frac{1}{4}\left(\frac{\Omega}{E_{LPB}(\beta)-E_X(\beta)}\right)^2}\right]^{-1};
\end{align}
are the excitonic and photonic fractions of the polariton, respectively. They were determined by fitting the theoretical dispersion curve (solid white line in the left panel of Figure \ref{Fig3}) to the experimental dispersion using the analytical expression for the lower polariton branch (LPB), which is the eigenvalue of the combination of excitonic states $E_X$ and photonic states $E_\gamma$:
\begin{equation}\label{eqLPB}
E_{LPB}(\beta) = \frac{1}{2}\left[E_X+E_\gamma(\beta)-\sqrt{\Omega^2+\left(E_X-E_\gamma(\beta)\right)^2}\right].
\end{equation}
In this equation $\Omega$ corresponds to the Rabi splitting, representative of the coupling strength between the photonic and excitonic modes, $E_X$ is the dispersion of the excitonic resonance, considered negligible in this work due to the large effective mass of excitons in GaN \cite{Vurgaftman_2001}, and $E_\gamma (\beta)$ is the dispersion relation of the fundamental $\mathrm{TE_0}$ mode propagating in the waveguide. This dispersion relation was calculated by solving the transcendental equation of the waveguide using the M\"uller-Traub method \cite{Dahlquist_2008}:
\begin{equation}
\left(1+i\frac{\kappa_1}{\alpha}\right) \left(1+i\frac{\kappa_2}{\alpha}\right) -\left(1-i\frac{\kappa_1}{\alpha}\right)\left(1-i\frac{\kappa_2}{\alpha}\right)e^{2i\alpha h}=0\, ;  
\end{equation}
where $\kappa_1$, $\alpha$, and $\kappa_2$ are the projections of the wavevector $\vec{k} = k_0 \vec{u}$ associated with the guided mode onto the ($Ox$) axis in air, GaN, and AlGaN, respectively; they are expressed in terms of $\beta$ and $k_0$ as follows: $\beta^2 - \kappa_1^2 = k_0^2$; $\alpha^2 + \beta^2 = n_{GaN}^2 k_0^2$; $\beta^2 - \kappa_2^2 = n_{AlGaN}^2 k_0^2$. In our simulations, we account for the material anisotropy and the dispersion of the refractive indices of GaN, $n_{GaN}$, and AlGaN, $n_{AlGaN}$. For the GaN index, we use an experimental index measured by ellipsometry from which the excitonic contributions have been subtracted, and the AlGaN index comes from reference \cite{Laws_2001}, taking into account the Al composition of the AlGaN alloy. The energy of the GaN excitonic transition $E_X$ = 3494.8 meV was determined experimentally by photoluminescence and reflectivity in a previous work \cite{Mechin_2025}. Fitting equation \ref{eqLPB} to the experimental results of Figure \ref{Fig3} yields a Rabi splitting value of $\Omega$ = 70 meV, while the broadening of the experimental dispersion gives an uncertainty of $\pm$15~meV in the Rabi determination.\par
\begin{figure}
	\includegraphics[width=\columnwidth]{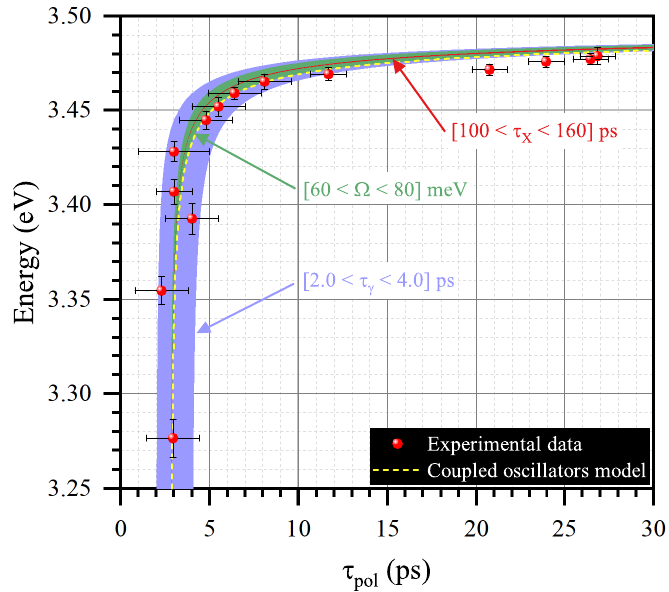}
	\caption{\label{Fig5} The red circles correspond to the polariton lifetimes experimentally determined as a function of the energy of the studied polariton state. The blue, green, and red regions correspond to the theoretical calculation of the polariton lifetime using equation \ref{eqtpol} and varying the photon lifetime between 2.0 ps and 4.0 ps, the Rabi splitting between 60 and 80 meV, and the excitonic reservoir decay time between 100 and 160 ps, respectively. The yellow dashed line corresponds to the polariton lifetime adjusted to the experimental results, which allows determining the following parameters: $\Omega$ = 80 meV, $\tau_X$ = 130 ps, $\tau_\gamma$ = 2.8 ps.}
\end{figure}
Photon lifetime in the waveguide $\tau_{\gamma}$ was determined by fitting the measured polariton lifetimes with equation \ref{eqtpol} and using the previously extracted Rabi splitting $\Omega$ and the exciton lifetime $\tau_X$. Figure \ref{Fig5} shows the results obtained with the two coupled oscillators model compared to the experimental results (red circles). Uncertainties of experimental data were estimated by (i) measuring the spectral full width at half maximum (FWHM) of the temporally analyzed signal of the polaritonic dispersion curve for the energy axis and (ii) varying the polariton lifetime used to do the fit of the temporal decays of polariton states along the dispersion curve for the time axis. The blue, green, and red areas, although the red area is seen as a red line, correspond to the calculation of the polariton lifetime using equation \ref{eqtpol}, and varying (i) the photon lifetime $\tau_{\gamma}$ between 2.0 ps and 4.0 ps (blue region); (ii) the Rabi splitting between 60 meV and 80 meV (green region); and (iii) the excitonic reservoir decay time $\tau_{X}$ between 100 ps and 160 ps (red region). Unless varied, the parameters were set at: $\Omega$ = 70 meV - $\tau_{X}$ = 130 ps - $\tau_{\gamma}$ = 2.8 ps. It can be observed that the results of the theoretical calculation are very little dependent on the excitonic reservoir lifetime, which only affects energies very close to the excitonic resonance (in fact, the dependance on excitonic lifetime is so small that the red shaded region in Figure \ref{Fig5} seems to be rather a line). In contrast, the results are highly dependent on the photon lifetime, highlighting the large photonic fraction of some of the studied polaritons, which enables us to determine precisely the actual photonic lifetime in the waveguide. The best fit of the model to the experimental results, represented by the yellow dashed line in Figure \ref{Fig5}, gives a value for the Rabi splitting: $\Omega$~=~80~meV, consistent with the results obtained from the fit of the experimental polariton dispersion curve and in agreement with the results from the literature on similar structures \citep{Ciers_2017, Brimont_2020, Souissi_2022, Souissi_2024}. The photon lifetime in the waveguide is estimated to be $\tau_{\gamma}$ = 2.8 ps, larger than the value estimated by Ciers et al. (0.9 ps) for an AlInN-based waveguide \cite{Ciers_2017} and larger than the characteristic photonic lifetimes measured in UV planar microcavities (0.2 ps) \cite{Levrat_2010}, thanks to the good confinement of the optical mode by total internal reflection. The large lifetime of photons in our heterostructure indicates a record $Q$-factor in the UV of 16~000, which is attributed to the low roughness scattering and reduced below-bandgap absorption in the bottom AlGaN cladding. Finally, the excitonic reservoir lifetime used for this fit, $\tau_{X}$ = 130 ps, is slightly lower than the results obtained experimentally from the direct luminescence of free excitons (168 ps). While a detailed study of the excitonic lifetime extracted from the fitting of the polariton lifetime might be interesting to analyze deeper the relaxation dynamics along the LPB, its effect is negligible on the current study (see red line in Figure~\ref{Fig5}) and does not affect any of the conclusions herein.

\section{Additional polariton feedings mechanisms at low temperature (5.3 K)}

\begin{figure}
	\includegraphics[width=\columnwidth]{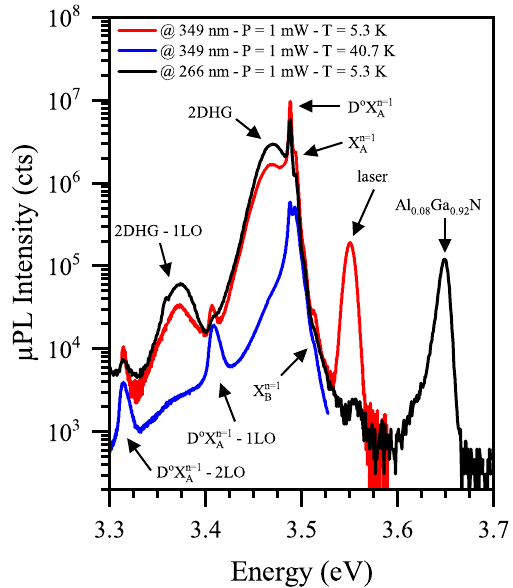}
	\caption{\label{Fig6} Low-excitation µPL spectra obtained with Ti:sapphire laser for different excitation wavelengths and work temperature. In blue, at 349 nm and T = 40.7 K. In red, at 349 nm and T = 5.3 K. In black, at 266 nm and T = 5.3 K. The decrease in temperature leads to relocalization of holes in the potential well associated to the 2DHG at the GaN/AlGaN top interface. The change in excitation wavelength (from 349 nm to 266 nm) allows to excite the AlGaN layer, as demonstrated by the presence of the AlGaN peak at 3.65 eV. The baseline of each spectrum is shifted vertically for clarity.}
\end{figure}
If the operation temperature is lowered to 5.3~K, as in reference \cite{Mechin_2025}, the thermal energy is no longer sufficient to delocalize excitons neither from neutral donors nor from the potential well created at the GaN/AlGaN interface due to the dielectric polarization mismatch across the interface. These changes modify the recombination dynamics of charge carriers in the structure, thereby modifying the dynamics of the excitonic reservoir and, consequently, the relaxation dynamics of polaritons along the dispersion curve.\par
Figure \ref{Fig6} shows the luminescence spectra obtained for different temperatures (5.3 K and 40.7 K) and different excitation wavelengths (349 nm and 266 nm). The measurements were performed at equal excitation power to facilitate the comparison. When the material is excited with a 349 nm laser at a temperature of 40.7 K (blue line), the excitonic transitions observed in GaN are $\mathrm{D^\circ X_A^{n=1}}$, $\mathrm{X_A^{n=1}}$, and $\mathrm{X_B^{n=1}}$, accompanied by one- and two-longitudinal optical phonon replicas, which are enhanced by the temperature. When the temperature decreases to 5.3 K (red line), a new, spectrally broad transition appears in the luminescence spectrum. This transition is associated with the recombination of holes confined in the two-dimensional hole gas (2DHG) at the interface between the GaN waveguide and the AlGaN cladding with electrons affected by the significant built-in electric field in the structure \cite{Mechin_2024}, which explains the spectral broadening of this transition. The appearance of this transition at low temperature can be explained by the reduction of the holes thermal energy which prevents them from escaping the shallow potential well associated with the 2DHG \cite{Jana_2016}. At higher energies, a signal resulting from the scattering of the laser excitation on the surface is observed. When the laser excitation wavelength is changed from 349 nm to 266 nm (black line), the photons that are not absorbed by the GaN waveguide become sufficiently energetic to excite the AlGaN layer, as evidenced in the luminescence spectrum by the peak at 3.65 eV. This new distribution of charge carrier also slightly modifies the intensity ratios observed between the GaN excitonic transitions and the 2DHG. Furthermore, when using the 266 nm excitation wavelength the ratio of the 2DHG to free exciton emission increases, indicating probably an additional recombination channel for the holes localized in the 2DHG. Indeed, these holes can now recombine with electrons photo-generated in the AlGaN cladding. Simultaneously, since the absorption of the laser excitation photons now occurs in the surface GaN and AlGaN layers, no luminescence from the thick bottom GaN layer is observed.\par
\begin{figure}
	\includegraphics[width=\columnwidth]{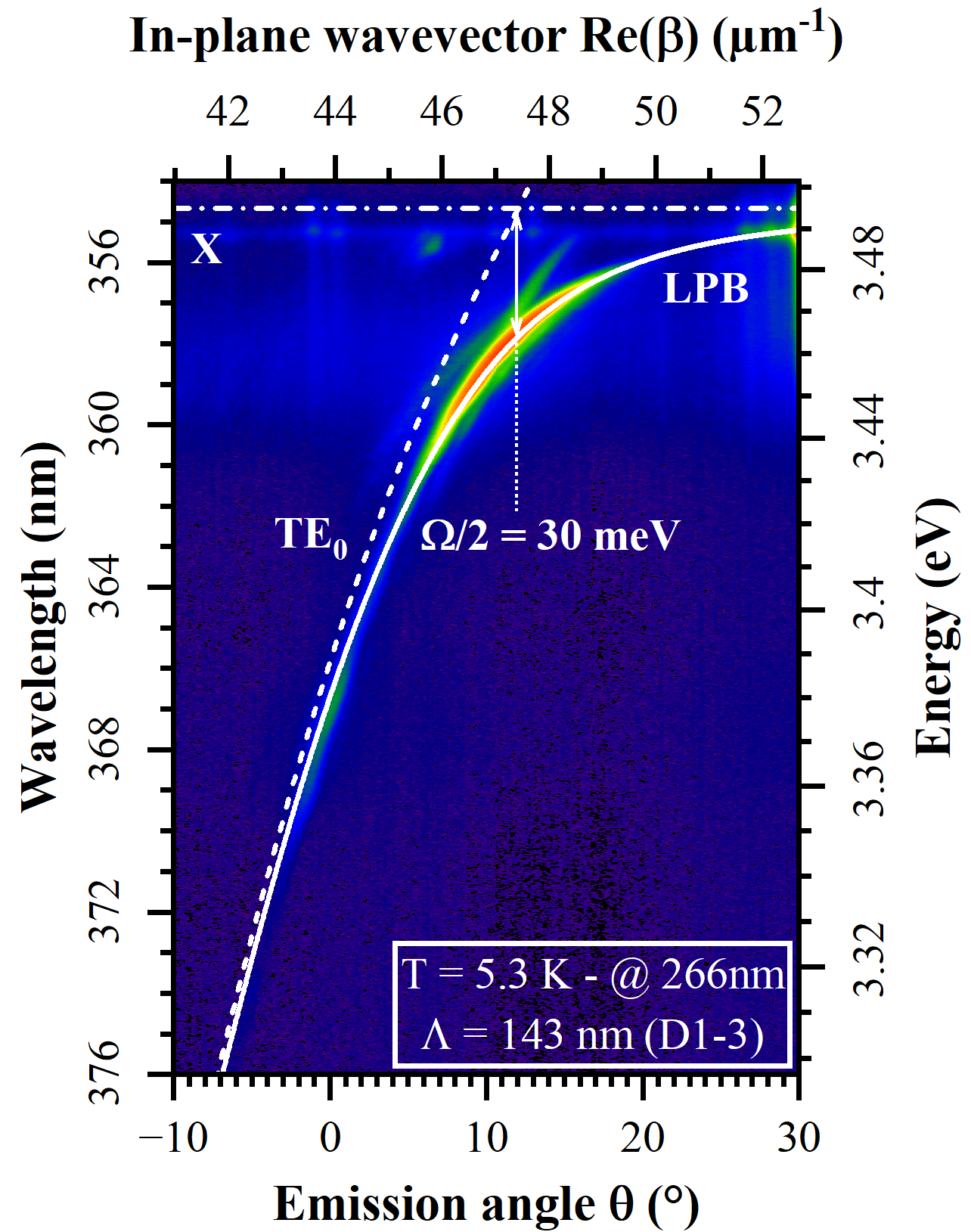}
	\caption{\label{Fig7} experimental measurement of the lower polariton branch (LPB) for a temperature of T = 5.3 K and a laser excitation power of 2 mW at 266 nm (the color scale corresponds to a logarithmic intensity scale). The solid white line represents the theoretical calculation of the polaritonic dispersion using a quasiparticle model. The dashed white lines correspond to the dispersion of the photonic mode propagating in the waveguide ($\mathrm{TE_0}$) and the excitonic resonance in the material at 3.4948 eV. The second mode observable above 3.46 eV, displaying a lower effective mass, corresponds to the $\mathrm{TM_0}$ mode strongly-coupled to GaN C exciton, as confirmed by polarization-resolved dispersion measurements.}
\end{figure}
To understand the effects of the new charge carrier spatial distribution on the polariton relaxation dynamics, we followed the same procedure for measuring the polariton lifetime as in the previous section. Figure \ref{Fig7} displays the lower polariton branch dispersion for an excitation power of 2 mW at a wavelength of 266 nm and a temperature of 5.3 K. The fit of the experimental dispersion using equation \ref{eqLPB} led to the determination of a Rabi splitting of $\Omega$~=~60~meV, slightly lower than the value determined for measurements at 40.7 K, which is probably due to a reduced exciton oscillator strength caused by  the built-in electric field.\par
\begin{figure}[h!]
	\includegraphics[width=\columnwidth]{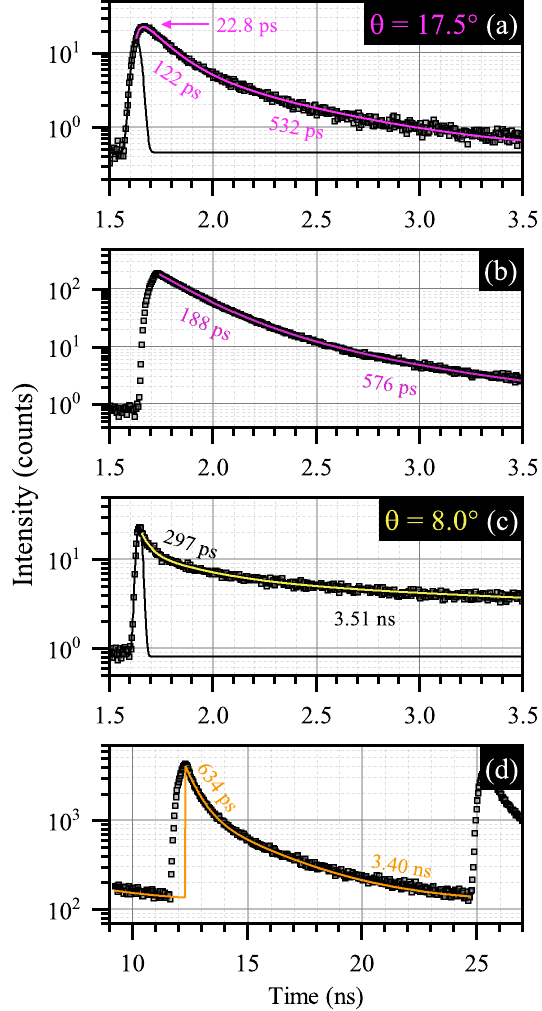}
	\caption{\label{Fig8} Time-resolved photoluminescence (TRPL) measurements realized at cryogenic temperature (5.3 K) and with an excitation wavelength set at 266 nm for : (a) the state of lower polaritonic branch with an angle of emission of 17.5° (P=2 mW); (b) the direct luminescence of the AlGaN transition (P=500 µW); (c) the state of lower polaritonic branch with an angle of emission of 8.0° (P=2 mW); (d) the direct luminescence of the 2DHG transition (P=2.5 mW). The gray squares represent the experimental data and the colored lines correspond to the fitting of experimental data using equation \ref{eqAlGaN} (Fig.\ref{Fig8}\textcolor{blue}{(a)}) and biexponential decays (Fig.\ref{Fig8}\textcolor{blue}{(b,c,d)}).}
\end{figure}

TRPL measurements along the polariton dispersion curve are presented in Figure \ref{Fig8} for two different angles, 17.5° and 8.0°, typical of the observed spectra. Above the energy of the 2DHG (3.4720 eV), a significant rise time accompanied by a long decay time is observed, as shown in Figure \ref{Fig8}\textcolor{blue}{(a)}. This decay time is much longer than the excitonic reservoir decay time expected at 5.3~K from previous measurements (24 ps at 1 mW~\cite{Mechin_2024}), indicating a modification of the mechanisms feeding polaritons at this LPB state. When the sample is excited with a wavelength of 266 nm, photons that are not absorbed by the GaN guiding layer can create carriers in the AlGaN cladding just below. The radiative recombination in AlGaN produces photons with enough energy to re-excite the GaN guiding layer, thereby giving rise to a second feeding mechanism for the excitonic reservoir. On the one hand, there is a direct and fast feeding of the excitonic reservoir immediately following the laser pulse, which decays rapidly due to the presence of a strong internal electric field~\cite{Mechin_2025}. Besides, a secondary and quasi-continuous feeding mechanism arises from photons generated via radiative recombination in the AlGaN layer. This second contribution is interpreted as being responsible for the long decay time observed in the TRPL measurements of LPB states above the energy of the 2DHG. Therefore, the measured excitonic reservoir lifetime should not be understood as the intrinsic exciton lifetime, but rather as a reflection of the reservoir feeding by the secondary process. The direct feeding mechanism, which decreases rapidly, may be masked by the contribution of the polariton emission. As a first approximation, photon recycling from the AlGaN layer can thus be considered the main feeding mechanism for the excitonic reservoir observed in the experimental results, and the temporal evolution of the polariton population can be described by the following differential equation:
\begin{align}
\frac{dn_{pol} (\beta, t)}{dt}&=C n_s(t) - \frac{n_{pol}(\beta, t)}{\tau_{pol}(\beta)} \quad \mathrm{with} ~ C\in [0, 1]\\
\mathrm{and} \quad n_s (t)&= C_1 \exp\left\lbrace-\frac{t}{\tau_{s1}}\right\rbrace + C_2 \exp\left\lbrace-\frac{t}{\tau_{s2}}\right\rbrace;
\end{align}
where $C$ is the fraction of photons emitted by the AlGaN layer that feed the GaN excitonic reservoir, and $n_s$ represents the density of photogenerated carriers in the AlGaN layer. $\tau_{s1}$ and $\tau_{s2}$ are the characteristic decay times of AlGaN luminescence, which have been experimentally measured using TRPL on AlGaN luminescence (Fig.\ref{Fig8}\textcolor{blue}{(b)}). The analytical expression describing the temporal evolution of the polariton density in the material can be written as:
\begin{align}\label{eqAlGaN}
n_{pol}(\beta, t)=&~a \times \exp\left\lbrace-\frac{t}{\tau_{s1}}\right\rbrace + b\times \exp\left\lbrace-\frac{t}{\tau_{s2}}\right\rbrace \nonumber \\
&- c\times\exp\left\lbrace-\frac{t}{\tau_{pol}(\beta)}\right\rbrace,
\end{align}
where $a$, $b$ and $c$ are positive constants. The fit of the experimental results in Figure \ref{Fig8}\textcolor{blue}{(a)} using Equation \ref{eqAlGaN} yields the following decay times: $\tau_{pol}$~=~22.8~ps, $\tau_{s1}$~=~122~ps, $\tau_{s2}$~=~532~ps. These results are consistent with the theoretical polariton lifetime calculation at 40~K (red dashed line in Figure \ref{Fig5}), which predicts $\tau_{pol}$~=~21.0~ps at the energy corresponding to an emission angle $\theta$~=~17.5°. Additionally, they align with the fit of the TRPL experimental results on AlGaN luminescence using a biexponential decay model (Fig.\ref{Fig8}\textcolor{blue}{(b)}), which gives $\tau_{s1}$~=~188~ps and $\tau_{s2}$~=~576~ps.\par
When measuring the temporal decay of polaritonic states at energies equal to or lower than the emission energy of the 2DHG, a biexponential decay is observed, with a component longer than those observed in LPB states at energies higher than the 2DHG emission energy. Fitting the experimental decay with a biexponential model (Fig. \ref{Fig8}\textcolor{blue}{(c)}) yields the following characteristic times: $\tau_{1}$ = 297 ps, $\tau_{2}$ = 3.51 ns. In GaN, decay times of this magnitude are usually obtained in the presence of an optical transition influenced by electric field effects. The only transition known in this sample with such a long decay time is the one associated with the 2DHG, involving holes confined in the potential well at the GaN/AlGaN interface and electrons spatially separated from the holes under the action of the built-in electric field \cite{Mechin_2024}. To confirm this hypothesis, TRPL measurements of the 2DHG luminescence were performed. A biexponential fitting of the experimental decay (Fig. \ref{Fig8}\textcolor{blue}{(d)}) yields similar times to those obtained from the polaritonic dispersion curve: $\tau_{1}$ = 634 ps, $\tau_{2}$ = 3.40 ns. We hypothesize that the 2DHG resonantly feeds the guided mode propagating in the GaN surface layer, explaining the long decay time observed in the TRPL measurements of the dispersion curve at low energy.

\section{Conclusion}
In this work we proposed a new experimental method to directly measure the lifetime of polaritonic states along their dispersion curve. Compared to existing methods reported in the literature, this technique does not require the measurement of the propagation length nor propagation time, which are both challenging to measure with high enough precision, nor the calculation or measurement of the group velocity. We perform these measurements on a GaN waveguide with extraction gratings deposited on its surface, enabling the outcoupling of the guided mode. By combining Fourier imaging with spatial filtering, each propagation constant can be directed to the entrance slit of the streak camera. A micrometric screw on the collection lens allows vertical translation of the Fourier plane, thereby adressing a different state of the lower polariton branch.\par
The main measurements of the polariton lifetime were carried out at a temperature of 40.9 K, with an excitation wavelength of 349 nm and a power of 10 mW. These experimental conditions were chosen to prevent the effects of the built-in electric field on the excitonic reservoir and to ensure direct feeding of the reservoir by photogenerated charge carriers. The experimental results reveal a polaritonic lifetime varying from about 30~ps, close to the excitonic reservoir energy, to about only some ps, revealing actually the lifetime of the photonic mode confined by total internal reflection ($\sim$~2.8 ps).\par
Finally, we also studied the sample close to liquid helium temperature (5.3 K) with a lower excitation wavelength to assess the effects of the built-in electric field and the photon recycling in the lower AlGaN cladding on the guided polaritons. These measurements revealed two other mechanisms of polariton feeding: (i) a secondary feeding of the excitonic reservoir through the recycling of photons emitted by radiative recombinations of electron-hole pairs in the lower AlGaN cladding, and (ii) a resonant excitation of the dispersion curve with photons emitted by the transition involving the 2DHG at the interface between the GaN waveguide and the AlGaN cladding.\par
In conclusion we present a new experimental technique to measure the polariton lifetime that we employ on a GaN based waveguide. By combining a fitting of the intensity decay at various wavevector and a coupled oscillators model, we determine a photon lifetime of $3 \pm 1$~ps corresponding to a record $Q$-factor in UV polaritonics of 16~000. We also demonstrate that the polariton lifetime is largely affected by excitonic properties and thus by the electric field. We conclude that although GaN based waveguides should be a very efficient platform for polariton lasing, care must be taken to prevent degrading the polariton lifetime.
\section{Acknowledgements}

The authors acknowledge fundings from the French National Research Agency (ANR-21-CE24-0019-NEWAVE). We also thank C2N, member of RENATECH, the French national network of large micro-nanofabrication facilities, for technological processes on our samples. We acknowledge support from GANEXT (ANR-11-LABX-0014); GANEXT belongs to the publicly funded ‘Investissements d’Avenir’ program managed by the Agence Nationale de la Recherche (ANR), France. Blandine Alloing and Jes\'us Z\'u\~niga-Pérez acknowledge financial support from the European Union under EIC Pathfinder open project "Polariton Neuromorphic Accelerator" (PolArt, Id: 101130304).



\begin{thebibliography}{9}
	\bibitem{Hopfield_1958} J. J. Hopfield, Theory of the Contribution of Excitons to the Complex Dielectric Constant of Crystals, Phys. Rev. 112, 1555 (1958).
	
	\bibitem{Agranovitch_1959} V. M. Agranovich, Zh. Eksp. Teor. Fiz. 37, 1555 (1959).
	
	\bibitem{Weisbuch_1992} C. Weisbuch, M. Nishioka, A. Ishikawa, and Y. Arakawa, Observation of the Coupled Exciton-Photon Mode Splitting in a Semiconductor Quantum Microcavity, Phys. Rev. Lett. 69, 3314 (1992).
	
	\bibitem{Walker_2013} P. M. Walker, L. Tinkler, M. Durska, D. M. Whittaker, I. J. Luxmoore, B. Royall, D. N. Krizhanovskii, M. S. Skolnick, I. Farrer, and D. A. Ritchie, Exciton Polaritons in Semiconductor Waveguides, Applied Physics Letters 102, 012109 (2013).
	
	\bibitem{Kasprzak_2006} J. Kasprzak et al., Bose–Einstein Condensation of Exciton Polaritons, Nature 443, 409 (2006).
	
	\bibitem{Kasprzak_2008} J. Kasprzak, D. D. Solnyshkov, R. André, L. S. Dang, and G. Malpuech, Formation of an Exciton Polariton Condensate: Thermodynamic versus Kinetic Regimes, Phys. Rev. Lett. 101, 146404 (2008).
	
	\bibitem{Richard_2005} M. Richard, J. Kasprzak, R. André, R. Romestain, L. S. Dang, G. Malpuech, and A. Kavokin, Experimental Evidence for Nonequilibrium Bose Condensation of Exciton Polaritons, Phys. Rev. B 72, 201301 (2005).
	
	\bibitem{Sun_2017} Y. Sun, P. Wen, Y. Yoon, G. Liu, M. Steger, L. N. Pfeiffer, K. West, D. W. Snoke, and K. A. Nelson, Bose-Einstein Condensation of Long-Lifetime Polaritons in Thermal Equilibrium, Phys. Rev. Lett. 118, 016602 (2017).
	
	\bibitem{Sanvitto_2016} D. Sanvitto and S. Kéna-Cohen, The Road towards Polaritonic Devices, Nature Mater 15, 1061 (2016).
	
	\bibitem{Amo_2009} A. Amo, J. Lefrère, S. Pigeon, C. Adrados, C. Ciuti, I. Carusotto, R. Houdré, E. Giacobino, and A. Bramati, Superfluidity of Polaritons in Semiconductor Microcavities, Nature Phys 5, 805 (2009).
	
	\bibitem{Lagoudakis_2009} K. G. Lagoudakis, T. Ostatnický, A. V. Kavokin, Y. G. Rubo, R. André, and B. Deveaud-Plédran, Observation of Half-Quantum Vortices in an Exciton-Polariton Condensate, Science 326, 974 (2009).
	
	\bibitem{Lagoudakis_2008} K. G. Lagoudakis, M. Wouters, M. Richard, A. Baas, I. Carusotto, R. André, L. S. Dang, and B. Deveaud-Plédran, Quantized Vortices in an Exciton–Polariton Condensate, Nature Phys 4, 706 (2008).
	
	\bibitem{Hivet_2012} R. Hivet et al., Half-Solitons in a Polariton Quantum Fluid Behave like Magnetic Monopoles, Nature Phys 8, 724 (2012).
	
	\bibitem{Kavokin_2005} A. Kavokin, G. Malpuech, and M. Glazov, Optical Spin Hall Effect, Phys. Rev. Lett. 95, 136601 (2005).
	
	\bibitem{Leyder_2007} C. Leyder, M. Romanelli, J. P. Karr, E. Giacobino, T. C. H. Liew, M. M. Glazov, A. V. Kavokin, G. Malpuech, and A. Bramati, Observation of the Optical Spin Hall Effect, Nature Phys 3, 628 (2007).
	
	\bibitem{Savvidis_2000} P. G. Savvidis, J. J. Baumberg, R. M. Stevenson, M. S. Skolnick, D. M. Whittaker, and J. S. Roberts, Angle-Resonant Stimulated Polariton Amplifier, Phys. Rev. Lett. 84, 1547 (2000).
	
	\bibitem{Saba_2001} M. Saba et al., High-Temperature Ultrafast Polariton Parametric Amplification in Semiconductor Microcavities, Nature 414, 731 (2001).
	
	\bibitem{Stevenson_2000} R. M. Stevenson, V. N. Astratov, M. S. Skolnick, D. M. Whittaker, M. Emam-Ismail, A. I. Tartakovskii, P. G. Savvidis, J. J. Baumberg, and J. S. Roberts, Continuous Wave Observation of Massive Polariton Redistribution by Stimulated Scattering in Semiconductor Microcavities, Phys. Rev. Lett. 85, 3680 (2000).
	
	\bibitem{Walker_2015} P. M. Walker, L. Tinkler, D. V. Skryabin, A. Yulin, B. Royall, I. Farrer, D. A. Ritchie, M. S. Skolnick, and D. N. Krizhanovskii, Ultra-Low-Power Hybrid Light–Matter Solitons, Nat Commun 6, 8317 (2015).
	
	\bibitem{Imamoglu_1996} A. Imamog¯lu, R. J. Ram, S. Pau, and Y. Yamamoto, Nonequilibrium Condensates and Lasers without Inversion: Exciton-Polariton Lasers, Phys. Rev. A 53, 4250 (1996).
	
	\bibitem{Brimont_2020} C. Brimont et al., Strong Coupling of Exciton-Polaritons in a Bulk $\mathrm{Ga}\mathrm{N}$ Planar Waveguide: Quantifying the Coupling Strength, Phys. Rev. Appl. 14, 054060 (2020).
	
	\bibitem{Rosenberg_2016} I. Rosenberg, Y. Mazuz-Harpaz, R. Rapaport, K. West, and L. Pfeiffer, Electrically controlled mutual interactions of flying waveguide dipolaritons, Phys. Rev. B 93, 195151 (2016).
	
	\bibitem{Suarez_2020} D. G. Suárez-Forero et al., Electrically controlled waveguide polariton laser, Optica, OPTICA 7, 1579 (2020).
	
	\bibitem{Amo_2016} A. Amo, T. C. H. Liew, C. Adrados, R. Houdré, E. Giacobino, A. V. Kavokin, and A. Bramati, Exciton–polariton spin switches, Nature Photon 4, 361 (2010).
	
	\bibitem{Leblanc_2020} C. Leblanc, G. Malpuech, and D. D. Solnyshkov, High-frequency exciton-polariton clock generator, Phys. Rev. B 101, 115418 (2020).
	
	\bibitem{Zasedatelev_2019} A. V. Zasedatelev, A. V. Baranikov, D. Urbonas, F. Scafirimuto, U. Scherf, T. Stöferle, R. F. Mahrt, and P. G. Lagoudakis, A room-temperature organic polariton transistor, Nat. Photonics 13, 378 (2019).

	\bibitem{Ellenbogen_2011} T. Ellenbogen and K. B. Crozier, Exciton-polariton emission from organic semiconductor optical waveguides, Phys. Rev. B 84, 161304 (2011).
	
	\bibitem{Pirotta_2014} S. Pirotta, M. Patrini, M. Liscidini, M. Galli, G. Dacarro, G. Canazza, G. Guizzetti, D. Comoretto, and D. Bajoni, Strong coupling between excitons in organic semiconductors and Bloch surface waves, Applied Physics Letters 104, 051111 (2014).

	\bibitem{Jamadi_2017} O. Jamadi et al., Edge-emitting polariton laser and amplifier based on a ZnO waveguide, Light: Science \& Applications 7, (2017).
	
	\bibitem{Kreyder_2023} G. Kreyder, L. Hermet, P. Disseix, F. Médard, M. Mihailovic, F. Réveret, S. Bouchoule, C. Deparis, J. Zuñiga-Pérez, and J. Leymarie, Lasing in a ZnO waveguide: Clear evidence of polaritonic gain obtained by monitoring the continuous exciton screening, Phys. Rev. B 107, 125307 (2023).

	\bibitem{Christopoulos_2007} S. Christopoulos et al., Room-Temperature Polariton Lasing in Semiconductor Microcavities, Phys. Rev. Lett. 98, 126405 (2007).
	
	\bibitem{Christmann_2008} G. Christmann, R. Butté, E. Feltin, J.-F. Carlin, and N. Grandjean, Room temperature polariton lasing in a GaN/AlGaN multiple quantum well microcavity, Applied Physics Letters 93, 051102 (2008).

	\bibitem{Di-Paola_2021} D. M. Di Paola et al., Ultrafast-nonlinear ultraviolet pulse modulation in an AlInGaN polariton waveguide operating up to room temperature, Nat Commun 12, 3504 (2021).
	
	\bibitem{Amano_1989} H. Amano, M. Kito, K. Hiramatsu, and I. Akasaki, P-Type Conduction in Mg-Doped GaN Treated with Low-Energy Electron Beam Irradiation (LEEBI), Jpn. J. Appl. Phys. 28, L2112 (1989).

	\bibitem{Nakamura_1992} S. Nakamura, T. M. T. Mukai, and M. S. M. Senoh, Si- and Ge-Doped GaN Films Grown with GaN Buffer Layers, Jpn. J. Appl. Phys. 31, 2883 (1992).
	
	\bibitem{Souissi_2022} H. Souissi et al., Ridge Polariton Laser: Different from a Semiconductor Edge-Emitting Laser, Phys. Rev. Appl. 18, 044029 (2022).

	\bibitem{Tassone_1997} F. Tassone, C. Piermarocchi, V. Savona, A. Quattropani, and P. Schwendimann, Bottleneck effects in the relaxation and photoluminescence of microcavity polaritons, Phys. Rev. B 56, 7554 (1997).
	
	\bibitem{Tassone_1999} F. Tassone and Y. Yamamoto, Exciton-exciton scattering dynamics in a semiconductor microcavity and stimulated scattering into polaritons, Phys. Rev. B 59, 10830 (1999).
	
	\bibitem{Solnyshkov_2014} D. D. Solnyshkov, H. Terças, and G. Malpuech, Optical amplifier based on guided polaritons in GaN and ZnO, Applied Physics Letters 105, 231102 (2014).

	\bibitem{Souissi_2024} H. Souissi et al., Mode-locked waveguide polariton laser, Optica 11, 962 (2024).
	
	\bibitem{Ciers_2017} J. Ciers, J. G. Roch, J.-F. Carlin, G. Jacopin, R. Butté, and N. Grandjean, Propagating Polaritons in III-Nitride Slab Waveguides, Phys. Rev. Applied 7, 034019 (2017).

	\bibitem{Detchprohm_1992} T. Detchprohm, K. Hiramatsu, K. Itoh, and I. Akasaki, Relaxation process of the thermal strain in the GaN/$\alpha$-Al2O3 heterostructure and determination of the intrinsic lattice constants of GaN free from the strain, Jpn. j. Appl. Phys 31, L1454 (1992).

	\bibitem{Mechin_2024} L. Méchin, F. Médard, J. Leymarie, S. Bouchoule, J.-Y. Duboz, B. Alloing, J. Zuñiga-Pérez, and P. Disseix, Experimental demonstration of a two-dimensional hole gas in a GaN/AlGaN/GaN based heterostructure by optical spectroscopy, Phys. Rev. B 109, 125401 (2024).

	\bibitem{Mechin_2025} L. Méchin, F. Médard, J. Leymarie, S. Bouchoule, B. Alloing, J. Zuñiga-Pérez, and P. Disseix, Effects of the built-in electric field on free and bound excitons in a polar GaN/AlGaN/GaN heterostructure, Phys. Rev. B 111, 155306 (2025).

	\bibitem{Hauswald_2014} C. Hauswald et al., Origin of the nonradiative decay of bound excitons in GaN nanowires, Phys. Rev. B 90, 165304 (2014).
	\bibitem{Hauswald_2013} C. Hauswald, T. Flissikowski, T. Gotschke, R. Calarco, L. Geelhaar, H. T. Grahn, and O. Brandt, Coupling of exciton states as the origin of their biexponential decay dynamics in GaN nanowires, Phys. Rev. B 88, 075312 (2013).
	
	\bibitem{Vurgaftman_2001} I. Vurgaftman, J. R. Meyer, and L. R. Ram-Mohan, Band parameters for III–V compound semiconductors and their alloys, Journal of Applied Physics 89, 5815 (2001).

	\bibitem{Dahlquist_2008} G. Dahlquist and Å. Björck, Numerical Methods in Scientific Computing (Society for Industrial and Applied Mathematics, 2008).

	\bibitem{Laws_2001} G. M. Laws, E. C. Larkins, I. Harrison, C. Molloy, and D. Somerford, Improved refractive index formulas for the $\mathrm{Al_x Ga_{1-x}N}$ and $\mathrm{In_y Ga_{1-y}N}$ alloys, Journal of Applied Physics 89, 1108 (2001).

	\bibitem{Levrat_2010} J. Levrat, R. Butté, E. Feltin, J.-F. Carlin, N. Grandjean, D. Solnyshkov, and G. Malpuech, Condensation phase diagram of cavity polaritons in GaN-based microcavities: Experiment and theory, Phys. Rev. B 81, 125305 (2010).

	\bibitem{Jana_2016} D. Jana and T. K. Sharma, An unambiguous identification of 2D electron gas features in the photoluminescence spectrum of AlGaN/GaN heterostructures, J. Phys. D: Appl. Phys. 49, 265107 (2016).

\end{thebibliography}
\end{document}